\documentclass[10pt,twocolumn]{article}

\usepackage[letterpaper,top=0.64in,bottom=0.69in,left=0.68in,right=0.68in]{geometry}
\usepackage{microtype}
\usepackage{newtxtext,newtxmath}
\usepackage{booktabs}
\usepackage{array}
\usepackage{tabularx}
\usepackage{multirow}
\usepackage{colortbl}
\usepackage{amsmath}
\usepackage{enumitem}
\usepackage{xcolor}
\usepackage{graphicx}
\usepackage{tikz}
\usepackage{pgfplots}
\usepackage{natbib}
\usepackage[hidelinks]{hyperref}
\usepackage{url}
\usepackage{caption}
\usepackage{subcaption}
\usepackage{float}

\usetikzlibrary{arrows.meta,positioning,fit,calc,matrix}
\usepgfplotslibrary{fillbetween}
\pgfplotsset{compat=1.18}

\definecolor{StudyNavy}{HTML}{17324D}
\definecolor{StudyBlue}{HTML}{28688C}
\definecolor{StudyTeal}{HTML}{26847A}
\definecolor{StudyGold}{HTML}{B7791F}
\definecolor{StudyRed}{HTML}{A64B4B}
\definecolor{StudyInk}{HTML}{1D2731}
\definecolor{StudyMuted}{HTML}{63717D}
\definecolor{StudyGrid}{HTML}{D6DEE4}
\definecolor{StudyPaper}{HTML}{F3F6F8}

\hypersetup{
  pdftitle={Answer-Reconstruction Search Density: Measuring the Query and Source Work Compressed by Conversational Answers},
  pdfauthor={Benjamin Tannenbaum},
  colorlinks=true,
  linkcolor=StudyBlue,
  citecolor=StudyBlue,
  urlcolor=StudyBlue
}

\setlist[itemize]{leftmargin=*,topsep=2pt,itemsep=1.5pt,parsep=0pt}
\setlist[enumerate]{leftmargin=*,topsep=2pt,itemsep=1.5pt,parsep=0pt}
\makeatletter
\renewcommand\section{\@startsection{section}{1}{\z@}%
  {-1.55ex plus -.35ex minus -.15ex}{0.65ex plus .1ex}%
  {\normalfont\large\bfseries\color{StudyNavy}}}
\renewcommand\subsection{\@startsection{subsection}{2}{\z@}%
  {-1.15ex plus -.25ex minus -.1ex}{0.4ex plus .1ex}%
  {\normalfont\normalsize\bfseries\color{StudyInk}}}
\renewcommand{\@maketitle}{%
  \newpage
  \null
  \vskip -2.2em
  \begin{center}%
    {\LARGE\bfseries\color{StudyNavy}\@title\par}%
    \vskip 0.75em
    {\normalsize
      \begin{tabular}[t]{c}
        \@author
      \end{tabular}\par}%
  \end{center}%
  \vskip 0.9em}
\makeatother

\renewenvironment{abstract}
  {\small\noindent\textbf{\color{StudyNavy}Abstract.}\ }
  {\par\vspace{0.6em}}

\title{Answer-Reconstruction Search Density:\\
Measuring the Query and Source Work Compressed by Conversational Answers}

\author{
  Benjamin Tannenbaum\\
  Aiso, Tel Aviv, Israel\\
  \texttt{ben@getaiso.com}
}
\date{}

\newcommand{\sourceRows}{1,180}
\newcommand{\uniqueN}{808}

\newcommand{\eligibleN}{183}
\newcommand{\unitN}{1,994}
\newcommand{\publicN}{6}
\newcommand{\privateQueryN}{803}
\newcommand{\ARSD}{\ensuremath{\mathrm{ARSD}}}
\newcommand{\FSD}{\ensuremath{\mathrm{FSD}}}
\newcommand{\PD}{\ensuremath{\mathrm{PD}}}
\newcommand{\Ret}{\ensuremath{\mathrm{Ret}}}

\begin{document}
\maketitle

\begin{abstract}
Conversational systems can collapse a visible sequence of web queries, result inspections, and source comparisons into one synthesized answer. Existing retrieval metrics evaluate ranking, effort, or factual support, but they do not measure the minimum conventional search work represented by a completed answer. We define \emph{answer-reconstruction search density} (\ARSD): the minimum number of distinct query actions required, under a fixed and dated reconstruction policy, to support a target share of atomic retrievable answer units. A parallel page-density measure separates query compression from source compression.

We evaluate the construct in two stages. First, we compute an exact structural facet-cover diagnostic over \eligibleN{} information-seeking conversations from a proprietary, consent-governed corpus. Their \unitN{} retained answer units require a median of 3 lexical facets (IQR 2--4) to cover 80\% under the primary policy, with 3.25 units covered per selected facet at the median. Multi-turn conversations have 0.90 higher unadjusted mean facet density (bootstrap 95\% CI 0.32--1.50), but the difference attenuates to 0.22 ($-0.35$--0.78) after adjusting for retained answer-unit count. A Poisson model gives the same result: the multi-turn incidence-rate ratio falls from 1.32 (1.09--1.56) to 1.06 (0.89--1.25). Thus, answer volume, rather than dialogue depth alone, accounts for most of the observed association. The median is unchanged across answer-unit caps of 8, 12, and 16; rank correlations across adjacent similarity policies are 0.85--0.93.

Second, a separate public calibration uses \publicN{} synthetic tasks, 30 prespecified units, and 36 fixed queries. Median live-web $\ARSD_{80}$ is 1.5 queries, median $\PD_{80}$ is 2 pages, and the opening query covers 70\% of units. The pilot validates measurement feasibility, not a corpus-level web estimate. Search density is a policy-relative measure of informational compression, not correctness, persuasion, or user benefit.
\end{abstract}


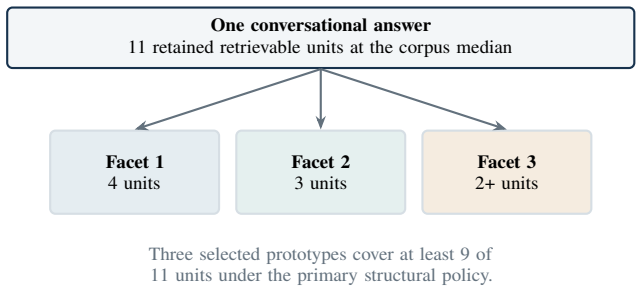
\begin{figure}[H]
\centering
\begin{tikzpicture}[
  font=\scriptsize,
  answer/.style={rounded corners=2pt,draw=StudyNavy,thick,fill=StudyPaper,
    align=center,text width=0.92\columnwidth,minimum height=8mm},
  facet/.style={rounded corners=2pt,draw=StudyGrid,thick,align=center,
    text width=20mm,minimum height=11mm},
  arrow/.style={-{Stealth[length=1.8mm]},thick,draw=StudyMuted}
]
\node[answer] (answer) {\textbf{One conversational answer}\\11 retained retrievable units at the corpus median};
\node[facet,fill=StudyTeal!12,below=8mm of answer.south] (q2)
  {\textbf{Facet 2}\\3 units};
\node[facet,fill=StudyBlue!12,left=2mm of q2] (q1)
  {\textbf{Facet 1}\\4 units};
\node[facet,fill=StudyGold!14,right=2mm of q2] (q3)
  {\textbf{Facet 3}\\2+ units};
\draw[arrow] (answer.south) -- (q1.north);
\draw[arrow] (answer.south) -- (q2.north);
\draw[arrow] (answer.south) -- (q3.north);
\node[below=3mm of q2,align=center,text=StudyMuted,text width=0.9\columnwidth]
  {Three selected prototypes cover at least 9 of 11 units under the primary structural policy.};
\end{tikzpicture}
\caption{The answer-first measurement perspective. The displayed values are corpus medians. Structural facets are not observed web queries; live query actions are measured separately.}
\label{fig:concept}
\end{figure}

\section{Introduction}

A conventional web search exposes its process. The user reformulates a query, opens results, compares sources, and decides when enough information has been collected. A conversational system can hide that sequence behind one response. A request for a product recommendation may yield suitability criteria, alternatives, constraints, prices, maintenance advice, and caveats in a single synthesized answer.

This shift creates a measurement gap. Query counts remain central to interactive information retrieval, search analytics, and marketing demand measurement, yet a single conversational turn can express and resolve several queryable aspects. Conversely, a long answer can repeat one idea. Neither prompt count, answer length, nor dialogue depth reveals how much conventional search work the delivered information represents.

The closest research traditions address different parts of the problem. Interactive IR measures effort and task success \citep{kelly2009methods,jiang2015satisfaction,shah2023task}. Conversational retrieval rewrites context-dependent turns or fans one need into several aspect queries \citep{dalton2020cast,kumar2020pipeline,abbasiantaeb2026multi}. Retrieval-augmented generation decomposes requests to balance evidence breadth against cost \citep{trivedi2023ircot,petcu2026decomposition}. Long-form factuality systems split answers into claims and search for support \citep{min2023factscore,gao2023citations,wei2024safe}. None of these lines reports the minimum \emph{distinct} conventional queries and source pages required to cover a fixed share of a completed answer.

We formalize that counterfactual as \emph{answer-reconstruction search density}. The longer name separates our construct from unrelated uses of ``search density'' in organizational and spatial search \citep{mazzelli2020density,kalbaugh1993density}. We use the short form only after this distinction.

The paper addresses four research questions:

\begin{description}[leftmargin=0pt,labelindent=0pt,style=nextline,itemsep=2pt]
  \item[\textbf{RQ1: Magnitude.}] How many distinct structural facets are present in observed information-seeking answers?
  \item[\textbf{RQ2: Robustness.}] How sensitive is that estimate to the target coverage, lexical similarity rule, and answer-unit cap?
  \item[\textbf{RQ3: Depth.}] Is multi-turn dialogue associated with density after accounting for the amount of answer material?
  \item[\textbf{RQ4: Operationalization.}] Can query density and page density be measured on a dated live-web snapshot?
\end{description}

Our contributions are: (1) an answer-first query and page-cover construct; (2) an exact partial set-cover formulation; (3) a full-population structural analysis with two-dimensional policy sensitivity, cap robustness, tail analysis, and adjusted count models; and (4) a fully public live-web calibration with claim-level support decisions. The empirical design deliberately separates the large private structural analysis from the small public web study.

\section{Related Work}

\subsection{Search tasks, sessions, and effort}

Search is an evolving process rather than an isolated query. Berrypicking describes how each finding changes the next information need \citep{bates1989berrypicking}; exploratory search emphasizes learning and investigation \citep{marchionini2006exploratory}; information-foraging theory models information gain under effort costs \citep{pirolli1999foraging}. Interactive IR accordingly evaluates action sequences, outcomes, satisfaction, and effort \citep{kelly2009methods,moffat2017expectations}. Naturalistic studies use query count, time, and reformulation as observable effort signals \citep{odijk2015struggling,vuong2019naturalistic}.

Search density reverses the direction of measurement. Rather than starting from observed actions and estimating effort or satisfaction, it starts from a delivered answer and asks for the smallest action set that could reconstruct a fixed share of its retrievable content. This is an answer-level search-work equivalent, not a claim about what a particular user actually did. Earlier database-interface research compared actual behavior with a task-specific minimum number of necessary queries \citep{jarke1985field}; our construct extends the minimum-action idea to open-web answer reconstruction and separates query actions from supporting pages.

\subsection{Conversational and multi-aspect retrieval}

Conversational information seeking models interactions in which context, clarification, and system initiative change retrieval \citep{radlinski2017framework,zamani2023cis}. CAsT and query-rewriting work recover the current turn's information need from dialogue history \citep{dalton2020cast,kumar2020pipeline}. Multi-aspect query generation shows that one rewrite may miss distinct aspects and that independent retrieval can improve coverage \citep{abbasiantaeb2026multi}. Query-decomposition methods in RAG similarly trade exploration against retrieval cost and noise \citep{petcu2026decomposition}.

These methods optimize retrieval \emph{before} answer generation. Search density evaluates a fixed answer \emph{after} generation. Turn count is not assumed to equal query count. Natural multi-turn prompts can be self-contained, and earlier assistant text is not always necessary for a high-quality continuation \citep{huang2026ownwords}. A follow-up may refine one facet, correct a detail, or open a new task.

\subsection{Atomic support, browsing, and diversity}

FActScore, ALCE, VeriScore, and $Q^2$ establish that long answers require claim-level support evaluation \citep{min2023factscore,gao2023citations,song2024veriscore,honovich2021q2}. SAFE is especially close: it decomposes a response into facts, issues web searches, and judges evidence \citep{wei2024safe}. Its queries are evaluator operations per fact; it does not minimize reusable query or page sets or interpret that minimum as conventional-search compression.

WebGPT and web-agent benchmarks record queries, page visits, and citations while producing answers \citep{nakano2021webgpt,song2025bearcubs}. They evaluate answer quality or agent success, not the minimum reconstruction set for a completed answer. Diversity metrics such as $\alpha$-nDCG reward nonredundant subtopic coverage within ranked results \citep{clarke2008novelty}. Search density instead minimizes actions across queries or pages at a fixed answer-unit coverage target.

\section{Formal Definition}

\subsection{Units, policies, and support}

For conversation $c$, let $A_c=\{a_1,\ldots,a_m\}$ be its atomic, externally retrievable answer units. Purely social, stylistic, and conversational material is excluded. A reconstruction policy $\pi$ fixes:

\begin{itemize}
  \item candidate-query generation;
  \item search provider, locale, date, and result depth;
  \item the support rule for a unit-page pair;
  \item the query budget and stopping rule.
\end{itemize}

Let $Q_c^\pi=\{q_1,\ldots,q_n\}$ be the candidate query actions and $M_{ij}=1$ when the evidence returned by query $q_j$ directly supports unit $a_i$. For target coverage $\tau\in(0,1]$, answer-reconstruction search density is the optimum of the partial set-cover problem

\begin{align}
\ARSD_\tau(c;\pi)
  = \min_{\mathbf{x},\mathbf{z}}\quad &\sum_{j=1}^{n}x_j \label{eq:arsd}\\
\text{s.t.}\quad
  &z_i \leq \sum_{j=1}^{n} M_{ij}x_j,\quad i=1,\ldots,m, \nonumber\\
  &\sum_{i=1}^{m}z_i \geq \lceil \tau m\rceil, \nonumber\\
  &x_j,z_i\in\{0,1\}. \nonumber
\end{align}

The same optimization over distinct supporting pages defines page density $\PD_\tau$. Maximum attainable coverage within the candidate budget defines retrievability:

\begin{equation}
\Ret(c;\pi)=\frac{1}{m}\left|\bigcup_{j=1}^{n}\{a_i:M_{ij}=1\}\right|.
\end{equation}

If $\Ret(c;\pi)<\tau$, $\ARSD_\tau$ is right-censored rather than assigned the maximum budget.

\subsection{Interpretation and basic properties}

The construct is policy-relative. For fixed $\pi$, $\ARSD_\tau$ is nondecreasing in $\tau$; adding a candidate query cannot increase the optimum; and a stronger support rule can only preserve or increase it. Query and page density need not agree. One broad query can expose several specialized pages, while one authoritative page can support units found through multiple candidate queries.

Minimum set cover is NP-hard \citep{chvatal1979setcover}. Our candidate pools are small enough for exact bitmask dynamic programming. The solver enumerates attainable support unions, retains the smallest query count for each union, and selects the minimum count reaching $\lceil\tau m\rceil$, breaking ties in favor of greater coverage.

Search density is not factual accuracy, source quality, response length, satisfaction, cognitive effort, or persuasion. A compact false answer can be highly retrievable; a correct local fact can be difficult to retrieve. These outcomes require separate measurement.

\subsection{Structural facet-cover diagnostic}

Corpus-scale live retrieval is not required to study answer structure. We define a separate diagnostic, structural facet density $\FSD_{\tau,\gamma}$. Each answer unit is treated as a candidate prototype. TF-IDF cosine similarity defines

\begin{equation}
M_{ij}^{\mathrm{struct}}=
\mathbf{1}\!\left[i=j\ \lor\ \cos_{\mathrm{tfidf}}(a_i,a_j)\geq\gamma\right].
\end{equation}

Substituting $M^{\mathrm{struct}}$ into Eq.~\ref{eq:arsd} yields the minimum prototype set covering $\tau$ of an answer's units. $\FSD$ measures lexical facet fragmentation. It is \emph{not} an observed query count and not a formal lower bound on live-web $\ARSD$: one broad search can retrieve lexically diverse units, while semantically equivalent units can use different words. We report the two quantities under distinct labels throughout.

\section{Data and Governance}

\subsection{Consent-governed corpus}

The private source is a proprietary, consent-governed collection of commercially relevant human-LLM conversations retained under applicable research and product-analysis agreements. The analyzed Google Drive snapshot contains \sourceRows{} populated rows. Identifier deduplication retains the longest record per identifier; exact normalized transcript deduplication then leaves \uniqueN{} observed conversations.

No organization identity is used as a feature, label, group, example, table row, or figure annotation. Raw transcripts, record identifiers, organization names, and private query strings are excluded from the paper and arXiv package. Illustrative wording is synthetic or de-identified and is never attributed to a source organization.

\begin{table}[t]
\centering
\small
\setlength{\tabcolsep}{4.2pt}
\begin{tabular}{@{}lrr@{}}
\toprule
Analysis stage & $N$ & Share of prior \\
\midrule
Populated source rows & 1,180 & -- \\
Unique observed conversations & 808 & 68.5\% \\
English-labeled conversations & 401 & 49.6\% \\
Eligible information-seeking cases & 183 & 45.6\% \\
\bottomrule
\end{tabular}
\caption{Corpus flow. Shares use the immediately preceding row. The 183 eligible cases form the full structural analysis population.}
\label{tab:flow}
\end{table}

\subsection{Eligibility and unit extraction}

The study retains conversations labeled English and informational, commercial, transactional, or navigational. Each must contain at least one parsed user turn, one assistant turn, and three retained answer units. Productive and creative tasks are excluded because reconstructing private rewriting or original composition through conventional web search is not a coherent counterfactual.

All assistant turns are processed. Markdown headings, greetings, closings, apologies, clarification invitations, and refusal boilerplate are removed. Numbered and bulleted items become candidates; remaining prose is sentence-split. A unit must contain 4--48 non-stopword terms. Candidate units with token-set Jaccard similarity at least 0.78 to an earlier unit are collapsed. If more than 16 remain, 16 positions are sampled evenly across the observed answer sequence so that later turns are not systematically discarded.

The resulting population contains \eligibleN{} conversations and \unitN{} retained units. It includes 97 single-turn and 86 multi-turn conversations; 166 are informational and 17 are commercial-action cases. The supplied language field is unknown for 351 of the 808 unique records, which is an important selection limitation.

\subsection{Public calibration}

The live-web calibration is deliberately independent. Eighteen synthetic public tasks were created without reference to a private conversation. Each task has one generic prompt, five answer units, one broad query, and five answer-aware unit queries. A deterministic six-task subset spans purchase, security, civic verification, organizational choice, and personal decision-making. The public artifact retains every task, query, support judgment, and source URL.

\section{Methods}

\subsection{Structural policy and estimands}

The primary structural policy sets answer-unit cap $K=16$, cosine threshold $\gamma=.15$, and target coverage $\tau=.80$. TF-IDF weights are estimated over the \unitN{} retained units. We solve the exact partial cover separately for every conversation.

RQ1 reports the median, interquartile range, mean with a 5,000-draw percentile-bootstrap confidence interval, the full distribution, and units covered per selected facet:

\begin{equation}
C_{80}(c)=\frac{\lceil .8|A_c|\rceil}{\FSD_{.8,.15}(c)}.
\end{equation}

RQ2 crosses $\tau\in\{.50,.80,1.00\}$ with $\gamma\in\{.10,.15,.20,.25\}$. It also repeats the full pipeline at caps $K\in\{8,12,16\}$, re-estimating document frequencies at each cap. Spearman rank correlations quantify whether policies preserve case ordering.

\subsection{Depth and task comparisons}

RQ3 first compares unadjusted means and the high-density tail $\mathbf{1}[\FSD_{.8,.15}\geq5]$ with 5,000-draw stratified-group bootstraps. It then fits two descriptive model families:

\begin{align}
\FSD_i &= \alpha+\beta D_i+\theta U_i+\varepsilon_i, \label{eq:ols}\\
\log E[\FSD_i] &= \alpha+\beta D_i+\theta U_i, \label{eq:pois}
\end{align}

where $D_i$ indicates multi-turn dialogue and $U_i$ is retained unit count. A full sensitivity specification adds $\log_2(1+\text{assistant words})$ and commercial-action intent. OLS effects are reported in facets; Poisson effects are exponentiated as incidence-rate ratios (IRRs). Percentile intervals use 5,000 OLS and 3,000 Poisson case-resampling draws. A functional-form-free check estimates an overlap-weighted contrast within exact retained-unit-count strata, with a stratified bootstrap and within-stratum permutation test. These are descriptive adjustments, not causal mediation models.

We also plot density within uncapped answer-size bins and complementary cumulative distributions by dialogue depth. Spearman correlations summarize associations with assistant words, user turns, retained units, and uncapped units.

\subsection{Live-web measurement}

On 20 July 2026, 36 fixed queries for the six public cases were submitted through a supported public web-search interface. A reviewer opened returned pages and assigned binary support only when a page directly substantiated a unit; lexical overlap alone was insufficient. Exact set cover over six query-support sets yields $\ARSD_{80}$; set cover over audited pages yields $\PD_{80}$. Opening coverage is the share supported by the broad query, and retrievability is coverage over all six queries.

An automated RSS pretest was rejected before scoring because compound technical queries repeatedly returned unrelated common-word senses. Its failure artifact is retained but contributes no reported result. A prepared private protocol contains 72 stratified cases and \privateQueryN{} privacy-minimized query candidates. The execution environment prohibited their external transmission, so no private query was sent and no corpus-level live-web value is reported.

\section{Results}

\subsection{RQ1: Answer facets are concentrated but heterogeneous}

The median eligible conversation contains 11 retained units (IQR 7--16). Before capping, the median is also 11, the IQR is 7--23, and the maximum is 142. Sixty-three cases (34.4\%) exceed the 16-unit cap.

At the primary policy, median structural facet density is 3 (IQR 2--4); the mean is 3.28 (bootstrap 95\% CI 2.99--3.59). Thirty-two conversations need one facet, while 34 require at least five and 14 require at least seven. The long tail reaches 12. The median coverage yield is 3.25 answer units per selected facet (IQR 2.00--4.33).

\begin{figure}[t]
\centering
\begin{tikzpicture}
\begin{axis}[
  width=0.96\columnwidth,
  height=48mm,
  ybar,
  bar width=5.2pt,
  xmin=0.4,xmax=12.6,
  ymin=0,ymax=50,
  xtick={1,...,12},
  xlabel={Structural facet density $\FSD_{80,.15}$},
  ylabel={Conversations},
  tick label style={font=\scriptsize},
  label style={font=\small},
  ymajorgrids=true,
  grid style={draw=StudyGrid},
  axis line style={draw=StudyMuted},
  fill=StudyBlue!75,
  draw=StudyBlue
]
\addplot+[ybar,fill=StudyBlue!75,draw=StudyBlue] coordinates {
  (1,32) (2,45) (3,34) (4,40) (5,12) (6,6)
  (7,5) (8,3) (9,2) (10,2) (11,1) (12,1)
};
\end{axis}
\end{tikzpicture}
\caption{Full-population structural facet-density distribution ($N=183$). The measure is an exact cover over lexical answer-unit prototypes, not live queries.}
\label{fig:distribution}
\end{figure}
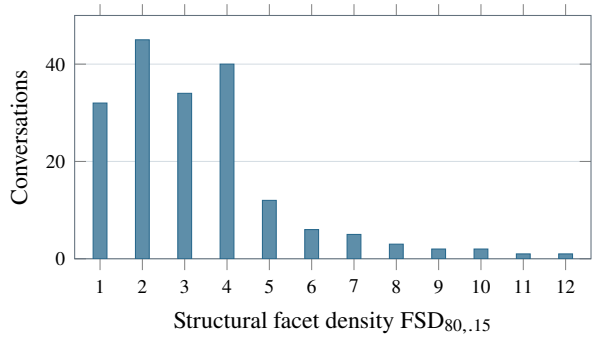

Length is informative but not equivalent to density. Spearman correlations with $\FSD_{80,.15}$ are 0.40 for assistant words, 0.22 for user turns, 0.44 for retained units, and 0.48 for uncapped units. Thus, answer volume accounts for part, but not all, of facet fragmentation.

\subsection{RQ2: Scale is policy-sensitive; ordering is stable}

Figure~\ref{fig:robustness} shows the full $4\times3$ policy grid. At 80\% coverage, the median rises from 2 to 5 as $\gamma$ moves from .10 to .25. At the primary $\gamma=.15$, it rises from 1 at 50\% coverage to 4 at full coverage. The magnitude is therefore inseparable from the stated policy.

Case ordering is more stable than the raw scale. Adjacent-threshold Spearman correlations are .85, .93, and .93. Answer-unit caps change the mean only from 3.09 at $K=8$ to 3.28 at $K=16$, while every cap yields median 3 and IQR 2--4. Rank correlations with the 16-unit policy are .89 at $K=8$ and .94 at $K=12$.

\begin{figure*}[t]
\centering
\begin{minipage}[t]{0.47\textwidth}
\centering
\small
\textbf{A. Median density across policy grid}\par\vspace{4pt}
\setlength{\tabcolsep}{8pt}
\renewcommand{\arraystretch}{1.38}
\begin{tabular}{@{}c|ccc@{}}
 & $\tau=.50$ & $\tau=.80$ & $\tau=1.00$ \\
\hline
$\gamma=.10$ & \cellcolor{StudyBlue!10}1 & \cellcolor{StudyBlue!20}2 & \cellcolor{StudyBlue!30}3 \\
$\gamma=.15$ & \cellcolor{StudyBlue!10}1 & \cellcolor{StudyBlue!30}\textbf{3} & \cellcolor{StudyBlue!42}4 \\
$\gamma=.20$ & \cellcolor{StudyBlue!20}2 & \cellcolor{StudyBlue!42}4 & \cellcolor{StudyBlue!54}5 \\
$\gamma=.25$ & \cellcolor{StudyBlue!20}2 & \cellcolor{StudyBlue!54}5 & \cellcolor{StudyBlue!66}6 \\
\end{tabular}
\end{minipage}
\hfill
\begin{minipage}[t]{0.47\textwidth}
\centering
\small
\textbf{B. Answer-unit cap sensitivity}\par\vspace{4pt}
\setlength{\tabcolsep}{6pt}
\renewcommand{\arraystretch}{1.27}
\begin{tabular}{@{}rrrrr@{}}
\toprule
Cap & Units & Median & Mean & $\rho$ vs.\ 16 \\
\midrule
8  & 1,289 & 3 & 3.09 & .888 \\
12 & 1,692 & 3 & 3.22 & .938 \\
16 & 1,994 & 3 & 3.28 & 1.000 \\
\bottomrule
\end{tabular}
\par\vspace{5pt}
\footnotesize Every cap has IQR 2--4. ``Units'' is the total retained after applying that cap to all 183 cases.
\end{minipage}
\caption{Robustness of structural facet density. Panel A varies target coverage and lexical similarity. The bold cell is the primary policy. Panel B re-estimates TF-IDF weights and exact covers under each unit cap.}
\label{fig:robustness}
\end{figure*}

\subsection{RQ3: The raw depth effect is mostly answer volume}

Multi-turn cases contain substantially more material: median uncapped unit count is 18 versus 8 for single-turn cases. Mean facet density is 3.76 versus 2.86, an unadjusted difference of 0.90 (95\% CI 0.32--1.50). Both groups have median density 3. The probability of density at least 5 is 24.4\% for multi-turn and 11.3\% for single-turn cases, a 13.1 percentage-point difference (2.0--24.2).

The association changes after accounting for answer size. In OLS, the multi-turn coefficient falls from 0.90 to 0.22 ($-0.35$--0.78) after adding retained unit count, and to 0.07 ($-0.51$--0.67) in the full model. The unit-count coefficient in the two-predictor model is 0.179 facets per retained unit (0.127--0.233).

The Poisson sensitivity model agrees. The unadjusted multi-turn IRR is 1.32 (1.09--1.56). It falls to 1.06 (0.89--1.25) after unit adjustment and 1.01 (0.85--1.20) in the full model. Each additional retained unit has an adjusted IRR of 1.060 (1.043--1.079). The exact unit-count analysis spans 174 cases in 12 overlapping strata and yields a 0.14-facet contrast ($-0.38$--0.64; within-stratum permutation $p=.69$).

\begin{figure*}[t]
\centering
\begin{subfigure}[t]{0.49\textwidth}
\centering
\begin{tikzpicture}
\begin{axis}[
  width=\textwidth,
  height=52mm,
  xmin=0.7,xmax=5.3,
  ymin=1.2,ymax=5.8,
  xtick={1,2,3,4,5},
  xticklabels={3--5,6--9,10--12,13--15,16+},
  xlabel={Uncapped answer-unit bin},
  ylabel={Mean structural facet density},
  tick label style={font=\scriptsize},
  label style={font=\small},
  ymajorgrids=true,
  grid style={draw=StudyGrid},
  axis line style={draw=StudyMuted},
  legend style={draw=none,fill=none,font=\scriptsize,at={(0.5,1.02)},anchor=south,legend columns=2}
]
\addplot+[StudyBlue,very thick,mark=*,error bars/.cd,y dir=both,y explicit] coordinates {
  (1,1.889) += (0,.333) -= (0,.333)
  (2,2.533) += (0,.435) -= (0,.433)
  (3,2.615) += (0,.692) -= (0,.692)
  (4,4.111) += (0,1.222) -= (0,1.111)
  (5,4.389) += (0,.833) -= (0,.667)
};
\addplot+[StudyTeal,very thick,mark=square*,error bars/.cd,y dir=both,y explicit] coordinates {
  (1,2.750) += (0,.750) -= (0,.875)
  (2,2.769) += (0,.769) -= (0,.769)
  (3,2.800) += (0,1.303) -= (0,1.000)
  (4,4.000) += (0,1.667) -= (0,1.667)
  (5,4.347) += (0,.755) -= (0,.735)
};
\legend{Single turn,Multi-turn}
\end{axis}
\end{tikzpicture}
\caption{Mean density within answer-size bins; bars are bootstrap 95\% intervals.}
\label{fig:bins}
\end{subfigure}
\hfill
\begin{subfigure}[t]{0.49\textwidth}
\centering
\begin{tikzpicture}
\begin{axis}[
  width=\textwidth,
  height=52mm,
  xmin=1,xmax=12,
  ymin=0,ymax=1.02,
  xtick={1,3,5,7,9,11},
  xlabel={Density threshold $d$},
  ylabel={Share with $\FSD_{80,.15}\geq d$},
  tick label style={font=\scriptsize},
  label style={font=\small},
  ymajorgrids=true,
  grid style={draw=StudyGrid},
  axis line style={draw=StudyMuted},
  legend style={draw=none,fill=none,font=\scriptsize,at={(0.5,1.02)},anchor=south,legend columns=2}
]
\addplot+[StudyBlue,very thick,mark=*,mark size=1.3pt] coordinates {
  (1,1) (2,.773) (3,.536) (4,.330) (5,.113) (6,.031)
  (7,.031) (8,.021) (9,.010) (10,.010)
};
\addplot+[StudyTeal,very thick,mark=square*,mark size=1.3pt] coordinates {
  (1,1) (2,.884) (3,.628) (4,.465) (5,.244) (6,.198)
  (7,.128) (8,.081) (9,.058) (10,.035) (11,.023) (12,.012)
};
\draw[StudyRed,dashed,thick] (axis cs:5,0) -- (axis cs:5,1);
\legend{Single turn,Multi-turn}
\end{axis}
\end{tikzpicture}
\caption{Complementary cumulative distribution; the dashed line marks the prespecified high-density tail.}
\label{fig:ccdf}
\end{subfigure}
\caption{Depth, answer volume, and the density tail. Multi-turn cases dominate the long-answer bin, but within-bin means are similar from 6 units upward.}
\label{fig:depth}
\end{figure*}
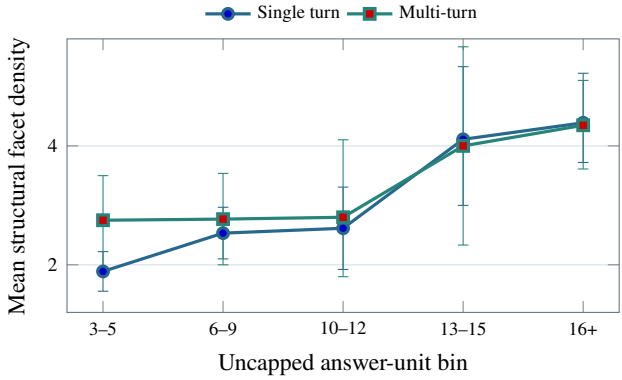
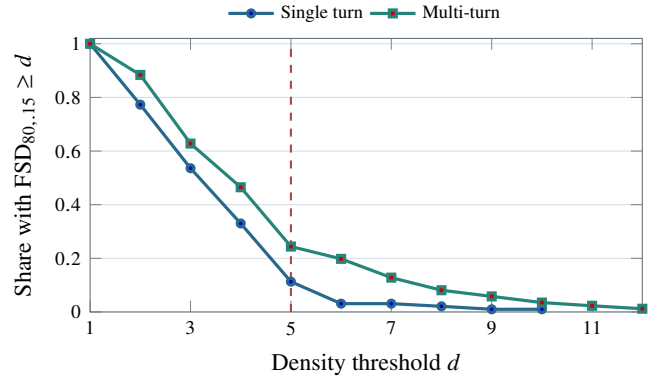

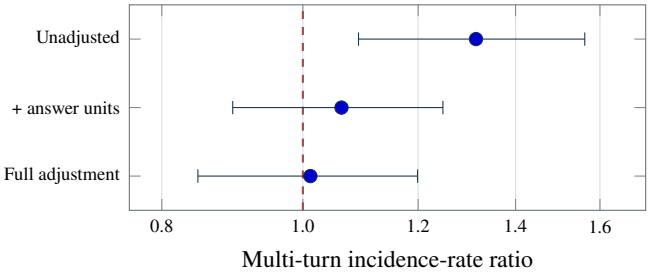
\begin{figure}[t]
\centering
\begin{tikzpicture}
\begin{axis}[
  width=0.96\columnwidth,
  height=43mm,
  xmode=log,
  xmin=.76,xmax=1.72,
  ymin=.5,ymax=3.5,
  ytick={1,2,3},
  yticklabels={Full adjustment,+ answer units,Unadjusted},
  xtick={0.8,1.0,1.2,1.4,1.6},
  xticklabels={0.8,1.0,1.2,1.4,1.6},
  xlabel={Multi-turn incidence-rate ratio},
  tick label style={font=\scriptsize},
  label style={font=\small},
  axis line style={draw=StudyMuted},
  xmajorgrids=true,
  grid style={draw=StudyGrid}
]
\draw[StudyRed,dashed,thick] (axis cs:1,.5) -- (axis cs:1,3.5);
\addplot+[only marks,mark=*,mark size=2.5pt,StudyNavy,
  error bars/.cd,x dir=both,x explicit] coordinates {
  (1.012,1) += (.187,0) -= (.165,0)
  (1.063,2) += (.185,0) -= (.168,0)
  (1.315,3) += (.247,0) -= (.223,0)
};
\end{axis}
\end{tikzpicture}
\caption{Poisson-model attenuation of the multi-turn association. Points are IRRs; bars are bootstrap 95\% intervals.}
\label{fig:forest}
\end{figure}

The binned analysis in Fig.~\ref{fig:bins} makes the same point without a functional-form assumption. Density rises sharply once answers reach 13 or more uncapped units, but depth-group means are nearly aligned within the 6--9, 10--12, 13--15, and 16+ bins. The tail difference in Fig.~\ref{fig:ccdf} is therefore real as a population description but should not be interpreted as an independent effect of conversational depth.

Commercial-action cases have mean density 3.53 versus 3.25 for informational cases. The difference is 0.28 facets with a wide interval ($-0.69$--1.47), and the high-density shares are nearly identical (17.6\% versus 17.5\%). The sample does not support a task-type difference.

\subsection{RQ4: Live query and page density diverge}

All \publicN{} public cases reach 80\% coverage within the six-query budget. Median manually audited $\ARSD_{80}$ is 1.5 queries (IQR 1--2); median $\PD_{80}$ is 2 pages (IQR 1--3). The broad opening query covers a median 70\% of the five units. Median retrievability is 100\%; one career-retraining case reaches exactly 80\%.

\begin{table}[t]
\centering
\footnotesize
\setlength{\tabcolsep}{3.3pt}
\begin{tabular}{@{}lcrrr@{}}
\toprule
Public task & Domain & Open & $\ARSD_{80}$ & $\PD_{80}$ \\
\midrule
Induction cooktop & Purchase & .60 & 2 & 3 \\
Heat pump & Purchase & .80 & 1 & 2 \\
Ransomware preparation & Security & 1.00 & 1 & 1 \\
Election-claim check & Civic & .60 & 2 & 3 \\
CRM selection & Organization & .80 & 1 & 1 \\
Career retraining & Decision & .60 & 2 & 2 \\
\bottomrule
\end{tabular}
\caption{Public manual calibration. Open is broad-query coverage. Every task and query is synthetic and independent of the private corpus.}
\label{tab:public}
\end{table}

\begin{figure}[t]
\centering
\begin{tikzpicture}
\begin{axis}[
  width=\columnwidth,
  height=50mm,
  ybar,
  bar width=4.4pt,
  ymin=0,ymax=3.45,
  ytick={0,1,2,3},
  ylabel={Actions at 80\% coverage},
  symbolic x coords={Cooktop,Heat pump,Ransomware,Election,CRM,Career},
  xtick=data,
  x tick label style={rotate=28,anchor=east,font=\scriptsize},
  tick label style={font=\scriptsize},
  label style={font=\small},
  ymajorgrids=true,
  grid style={draw=StudyGrid},
  axis line style={draw=StudyMuted},
  legend style={draw=none,fill=none,font=\scriptsize,at={(0.5,1.03)},anchor=south,legend columns=2}
]
\addplot+[fill=StudyBlue!76,draw=StudyBlue] coordinates {
  (Cooktop,2) (Heat pump,1) (Ransomware,1) (Election,2) (CRM,1) (Career,2)
};
\addplot+[fill=StudyTeal!70,draw=StudyTeal] coordinates {
  (Cooktop,3) (Heat pump,2) (Ransomware,1) (Election,3) (CRM,1) (Career,2)
};
\legend{Queries,Pages}
\end{axis}
\end{tikzpicture}
\caption{Query and page density in the public calibration. One broad query can expose evidence distributed across several source pages.}
\label{fig:public}
\end{figure}
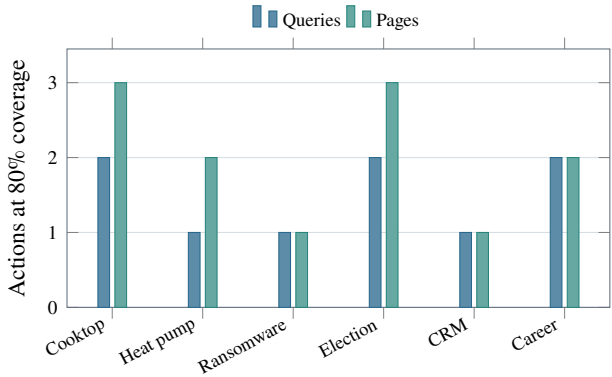

Three cases reach the target with one broad query. Only two of those also need one page; the heat-pump task needs two pages. This demonstrates why query compression is not source compression. It also shows why the structural facet diagnostic and live-web density are not interchangeable: a search result page can aggregate lexically diverse evidence.

\section{Discussion}

\subsection{The central empirical result}

The typical answer in this selected corpus contains 11 retained retrievable units organized into 3 lexical facets at the primary 80\% policy. That estimate is not reducible to word count, and the policy preserves case ordering reasonably well across similarity thresholds and unit caps.

The deeper result concerns conversation depth. Multi-turn cases are overrepresented in the high-density tail, but they also contain more than twice as many uncapped units at the median. Adjusted OLS, Poisson, and within-size-bin analyses all attenuate the depth association. The evidence supports a narrower conclusion than the unadjusted comparison: deeper dialogues often \emph{produce more answer material}, and that additional material carries more facets. The data do not show that the multi-turn form itself independently increases density.

\subsection{Implications for search and marketing}

Query-centric analytics observe expressions of demand, not the full information need. A synthesized answer can cover selection criteria, alternatives, constraints, availability, maintenance, and risk through one private interaction. Aggregate $\ARSD$ could translate conversation classes into policy-specific query-action equivalents, while $\PD$ estimates source fragmentation. Both can help explain why visible keyword volume may diverge from informational demand.

The unit of visibility also changes. A brand, policy position, or source can be absent from the opening prompt yet relevant to one reconstructed answer unit. Evaluation based on a single short keyword can therefore miss the source competition inside a synthesized response. Task-level bundles of answer units and evidence paths are a more faithful audit unit.

\subsection{Politics, security, and interface governance}

High-density answers concentrate several acts of question decomposition, source selection, and synthesis in one interface. Users may not see which subquestions were asked, which sources were omitted, or how uncertainty changed across facets. This concentration matters for security guidance and political information even when no purchase is involved.

Controlled experiments show that personalized LLM debates can influence opinions \citep{salvi2025persuasion}. Search density does not measure that effect and should not be treated as a persuasion score. It can instead serve as a sampling variable for later causal audits: high-density political or security answers are cases in which more informational selection is concentrated and may warrant stronger provenance, source-diversity, or manipulation tests.

\subsection{Design and evaluation}

Interfaces could expose a compact reconstruction trace: the number of answer units, searches, source pages, unsupported units, and major facets. High-density answers may justify expandable query trails or source-diversity warnings. For retrieval systems, lower density is not automatically better. A small query set can reduce latency and cost while also narrowing evidence. The relevant objective is adequate verified coverage under an explicit effort and diversity policy.

\section{Limitations}

First, the corpus is selected for commercial relevance and is not representative of all conversational-AI use. Only 183 conversations meet the English information-seeking criteria; 351 unique records have an unknown supplied language label, and the commercial-action group contains only 17 cases.

Second, unit extraction is deterministic and has not been independently double-coded. A sentence can contain multiple claims, a list item can be non-retrievable, and paraphrase deduplication can merge distinct details. The 16-unit cap affects 34.4\% of cases. Evenly spaced selection protects later turns but makes density conservative for long answers.

Third, $\FSD$ is lexical. It can split semantic paraphrases and merge units sharing vocabulary. The two-dimensional sensitivity analysis makes policy dependence visible but does not establish semantic validity. The structural result must not be reported as ``three web searches.''

Fourth, depth analyses are observational. Unit count can be both a consequence of additional turns and a proxy for answer complexity. Adjustment clarifies association but does not identify a causal pathway.

Fifth, the live-web calibration contains six synthetic cases and one reviewer. It establishes operational feasibility, not inter-rater reliability, population magnitude, or private-corpus validity. The queries are answer-aware, so their minimum understates unaided human search effort. Search results are also dynamic, localized, and provider-dependent.

Sixth, the approved private reconstruction could not be executed because the environment prohibited external transmission of corpus-derived strings. The paper therefore reports no private live-web estimate. The failed RSS pretest also shows that a technically complete retrieval run is not sufficient without face-validity and support audits.

Finally, density does not measure correctness, reading time, cognitive load, clicks, learning, decision quality, persuasion, or economic value. It should be combined with source quality and outcome measures in high-stakes settings.

\section{Conclusion}

Answer-reconstruction search density supplies an answer-level unit for a search process that conversational systems increasingly hide. It asks how many distinct queries and source pages are needed, under a documented policy, to recover a fixed share of a completed answer's retrievable content.

Across 183 consent-governed information-seeking conversations, the primary structural policy yields median facet density 3 and median coverage yield 3.25 units per selected facet. The headline depth difference is more nuanced than a raw comparison suggests: multi-turn cases have a larger high-density tail, but adjusted models and within-size-bin analyses show that answer volume accounts for most of the association. A separate six-case public calibration yields median live-web density 1.5 queries and median page density 2, demonstrating that query compression and source compression differ.

The next steps are independently coded unit and support audits, a larger multi-provider public benchmark, a privacy-compatible corpus reconstruction, and a human search experiment. Those stages are necessary before converting structural facets into claims about human effort or market-scale query displacement.

\section*{Ethics and Data Availability}

The private corpus is not released. Public materials include analysis code, aggregate and anonymized case-level outputs, synthetic tasks, public support decisions and URLs, LaTeX source, and the paper. They exclude raw transcripts, organization identities, record identifiers, private queries, and private search-result caches. No organization identity is used as a feature. The \privateQueryN{} prepared private candidates remained local and were not transmitted.

\bibliographystyle{plainnat}
\bibliography{references}

\appendix

\section{Exact-Cover Algorithm}

For each case, candidate support sets are encoded as bit masks over $m$ units. The exact solver proceeds as follows:

\begin{enumerate}
  \item Initialize a map containing the empty union with cost zero.
  \item For each candidate set, combine it with every currently reachable union.
  \item For each resulting union, retain the smallest number of selected candidates.
  \item Among unions covering at least $\lceil\tau m\rceil$ units, return the smallest cost, breaking ties by greater coverage.
\end{enumerate}

With at most 16 units and one prototype per unit in the structural analysis, the state space is at most $2^{16}$. The public calibration has five units and six queries. Exact rather than greedy solutions are therefore inexpensive and remove approximation error from the reported minima.

\section{Expanded Structural Sensitivity}

\begin{table}[H]
\centering
\footnotesize
\setlength{\tabcolsep}{3.8pt}
\begin{tabular}{@{}ccrrr@{}}
\toprule
$\gamma$ & Coverage & Median & Mean & IQR \\
\midrule
.10 & .50 & 1 & 1.30 & 1--2 \\
.10 & .80 & 2 & 2.37 & 1--3 \\
.10 & 1.00 & 3 & 3.63 & 2--5 \\
\addlinespace
.15 & .50 & 1 & 1.63 & 1--2 \\
.15 & .80 & 3 & 3.28 & 2--4 \\
.15 & 1.00 & 4 & 4.75 & 3--6 \\
\addlinespace
.20 & .50 & 2 & 2.07 & 1--2.5 \\
.20 & .80 & 4 & 4.28 & 2--5.5 \\
.20 & 1.00 & 5 & 5.93 & 3--8 \\
\addlinespace
.25 & .50 & 2 & 2.59 & 1--3 \\
.25 & .80 & 5 & 5.22 & 3--7 \\
.25 & 1.00 & 6 & 6.97 & 4--9.5 \\
\bottomrule
\end{tabular}
\caption{All 12 structural policy cells over the same 183 conversations.}
\label{tab:fullsens}
\end{table}

\section{Adjusted-Model Details}

\begin{table}[H]
\centering
\footnotesize
\setlength{\tabcolsep}{3.6pt}
\begin{tabular}{@{}lrrr@{}}
\toprule
Specification & Depth effect & 95\% CI & Fit \\
\midrule
OLS unadjusted & 0.900 & 0.319--1.498 & $R^2=.048$ \\
OLS + units & 0.221 & $-0.351$--0.782 & $R^2=.196$ \\
OLS full & 0.073 & $-0.514$--0.671 & $R^2=.209$ \\
\addlinespace
Poisson unadjusted & 1.315 & 1.093--1.562 & $R^2_M=.015$ \\
Poisson + units & 1.063 & 0.895--1.248 & $R^2_M=.064$ \\
Poisson full & 1.012 & 0.847--1.199 & $R^2_M=.068$ \\
\bottomrule
\end{tabular}
\caption{Multi-turn effects. OLS reports additive facets; Poisson reports IRRs. Full models add retained units, log assistant words, and commercial-action intent. $R^2_M$ is McFadden's pseudo-$R^2$.}
\label{tab:models}
\end{table}

\end{document}